# Beyond Accessibility: Lifting Perceptual Limitations for Everyone

**Michael Mauderer**
University of Dundee
Dundee, UK
m.mauderer@dundee.ac.uk

**Garreth W. Tigwell**
University of Dundee
Dundee, UK
g.w.tigwell@dundee.ac.uk

**Benjamin M. Gorman**
University of Dundee
Dundee, UK
b.gorman@dundee.ac.uk

**David R. Flatla**
University of Dundee
Dundee, UK
d.flatla@dundee.ac.uk

 

## Abstract
We propose that accessibility research can lay the foundation for technology that can be used to augment the perception of everyone. To show how this can be achieved, we present three case studies of our research in which we demonstrate our approaches for impaired colour vision, situational visual impairments and situational hearing impairment.

## Author Keywords
perception, augmentation, vision, hearing, accessibility

## Introduction
Accessibility research aims to create technologies that benefit people with disabilities. These technologies can range from prosthesis for amputees to digital tools to increase display contrast for people with low vision. Often the goal is to alleviate challenges brought on due to the impairment and achieve equality to people with typical abilities. In some cases, however, the tools that are created to deal with an impairment allow the people that use them to exceed typical abilities, or potentially provide a benefit to people of all abilities.

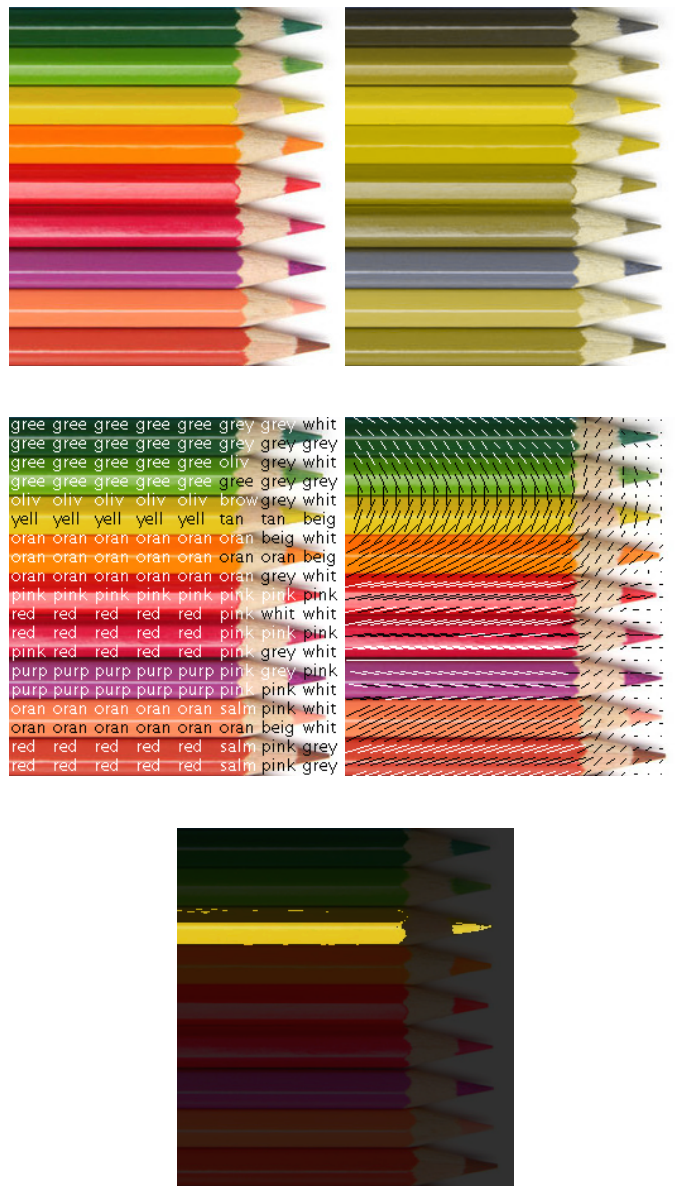

**Figure 1:** Examples of colour identification techniques for the task “select the yellow pencil”. Top left to bottom right: typical vision, impaired colour vision (ICV) simulation showing difficulty, our ColourNames, ColourMeters, and ColourPopper technique.

Neil Harbisson, who was born with total colour blindness, uses a device that records colour images in front of him and turn them into vibrations in his skull allowing him to ”hear“ the colours in the image. The device is set up in a way that allows him to not only see what is in front of him but also receive images from other sources, i.e., pictures sent to him from people in other locations [11], thus in a way expanding his perceptual capabilities beyond those of a typical person. An example of accessible technology that helps everyone are design guidelines for websites that address colours that are hard to differentiate for people with colour vision deficiencies and contrast for text which can be hard to read for people with low vision. While initially intended to alleviate challenges arising from impairments, these guidelines result in website designs that benefit everyone viewing the website through clearer colours and increased contrast [18].

The remainder of the paper will present digital assistive systems that deal with perceptual issues and show how from investigating (situational) disabilities we can develop technologies, tools, or design guidelines that can benefit and enhance everyone’s perception. To do this, we present three case studies that showcase our results for (1) impaired colour vision, (2) situational visual impairments, and (3) situational hearing impairment.

## Case Study 1: Impaired Colour Vision

Colour vision is a vital component of day-to-day living; its loss can severely limit a person’s livelihood (e.g., airplane pilots, electricians, dentists, and visual artists all require accurate colour vision), threaten health and safety (e.g., not spotting sunburns and rashes, misidentifying medication), make food preparation dangerous (e.g., undercooking meat, detecting mouldiness), and even challenge someone’s social acceptability (e.g., by inhibiting clothing coordination and home decorating) [3].

To help address the challenges imposed by impaired colour vision (ICV), we developed the concept of Situation-Specific Modelling (SSM) and used it to construct a model of colour perception tunable to an individual’s unique abilities and environment [5, 6, 7]. We also developed a recolouring tool for improving web colour differentiability for people with ICV that was designed to maintain the subjective experience of the original colour scheme (e.g., by keeping ‘warm’ colour schemes ‘warm’) while improving colour accessibility [8].

Next, we went beyond recolouring to focus on improving colour identification and discrimination. We developed revisualization techniques for mapping colour information to other visual properties (ColourID) [4], which improved colour identification to nearly 100% for participants with ICV (see Figure 1 for examples of the techniques). We also investigated the use of eye tracking based manipulations of colours to exploit with the eventual aim to make colours more differentiable to the observer [15].

Every technique we have developed has the characteristic of enabling people to perceive visual properties that they typically cannot (i.e., we make the invisible visible – in this case invisible *colour*). However, the same techniques can be used to let people with typical abilities ‘see the invisible’. Next, we plan to extend our techniques to give people the ability to perceive light outside of the visible spectrum: infrared and ultraviolet. We will apply the same techniques that work for people with ICV to enable all people to perceive information they normally wouldn’t be able to see at all. This has the potential to greatly improve health (e.g., by enabling rapid visual triage for fever in epidemic situations) and safety (e.g., giving airport security the ability to ‘see’ emitted heat).

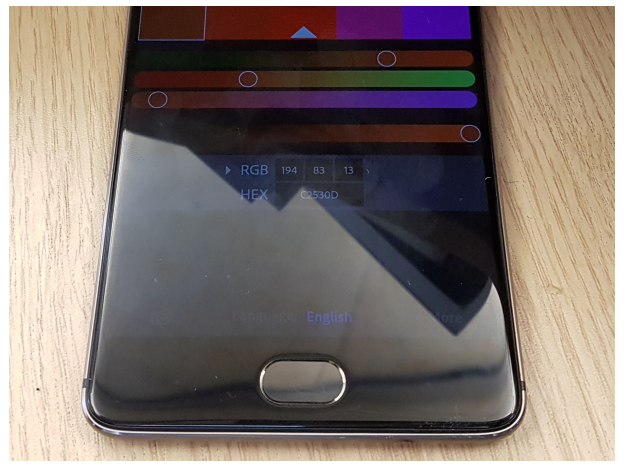

**Figure 2:** A smartphone in a brightly lit room, with half the device in shade. The low contrast menu text at the bottom of the screen becomes difficult to read.

| Amount of Information | Skill Type: ANALYTIC | HYBRID | SYNTHETIC |
|---|---|---|---|
| HIGH | | | |
| MED | | | |
| LOW | | | |

**Figure 3:** Our speechreading acquisition framework, with two dimensions: *Type of Skill* and *Amount of Information*, each split into three levels (*Analytic*/*Hybrid*/*Synthetic* and *Low*/*Medium*/*High*, respectively)

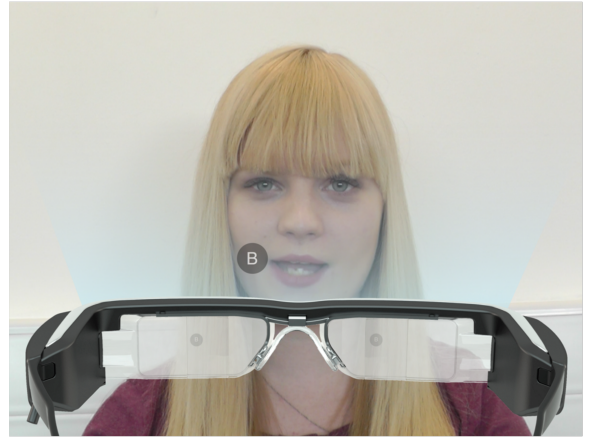

**Figure 4:** PhonemeViz mockup viewed through Epson Moverio glasses for the word "bat".

## Case Study 2: Situational Visual Impairment

A situational impairment is when a person experiences difficulty completing tasks that would not typically present problems (e.g., the difference when typing a message while stationary vs. running, or reading a message on a display of a mobile device on an overcast vs. sunny day). Various factors can cause a situational impairment, and it is not limited to environmental conditions. This concept within computing has been discussed by Newell [16] and Sears et al. [17].

The portability and computing power of mobile devices has extended our freedom for completing tasks at our convenience. However, with this freedom there are substantial issues; situational contexts can become difficult to control and can also be unpredictable. We focus specifically on situational visual impairment (SVI) when using mobile devices, which describes situations in which it becomes difficult to see content on the mobile device when it would otherwise not be an issue (e.g., bright light reflecting on the display and low contrast text as demonstrated in Figure 2).

Mobile screens are used in diverse situations and for diverse purposes, e.g., car navigation, in the medical industry, and education. Currently, we are looking at the SVIs experienced when using mobile devices under bright lighting. Although it is well documented that mobile screens typically degrade in quality as ambient light increase (e.g., [13, 14]), we do not know what tasks people are doing on their mobile devices when experiencing SVI, what other factors are involved that can exacerbate the problem, and what solutions people currently employ to overcome the problem.

We predict that there are different solutions required to reduce the occurrence of SVI since it is likely to involve multiple factors at any one time. For example, if the content being viewed has a low contrast design then it will become even harder to read on a display in direct sunlight. One solution would be to provide designers with a tool that helps them assess the robustness of their design's perceptibility under increasing ambient light, similar to an accessibility tool that we have previously developed to allow web designers to select a colour scheme for people with ICV, while maintaining the designers' creative freedom [19].

Solving this issue of situational visual impairment would be a benefit for everyone who is using mobile devices where overly bright conditions occur.

## Case Study 3: Situational Hearing Impairment

The WHO estimates that on a global scale 360 million people (~5%) worldwide have disabling hearing loss[1] [21]. Hearing loss is an everyday problem as it interferes with conversations by making it difficult to understanding what others are saying [20].

People who are deaf or have a hearing loss find that speechreading (SR) (commonly called lipreading) can overcome many of the barriers when communicating with others [2]. SR helps to improve conversational confidence (thereby reducing social isolation), enhance employability, and improve educational outcomes [20].

However, SR is a skill that takes considerable practice and training to acquire [12]. One reason for this is caused by nearly identical mouth shapes (visemes) producing several speech sounds (phonemes); there is not a one-to-many mapping from visemes to phonemes. This decreases comprehension and causes confusion and frustration during conversations. Publicly-funded SR classes are sometimes provided and have been shown to improve SR acquisition [1]. However, classes are only provided in few countries around the world; there is an insufficient number of classes running in areas in which they are provided (e.g., only 50 of an estimated 325 required classes are currently running in Scotland [1]) and classes require mobility to attend.

[1] Hearing loss of more than 40 dB in the better hearing ear in adults and more than 30 dB in the better hearing ear in children.

To help expand SR training, we developed a novel framework [10](Figure 3) that can be used to develop Speechreading Acquisition Tools (SATs) – a new type of technology designed specifically to improve SR acquisition. Through the development and release of SATs, people with hearing loss will be able to augment their class-based learning or learn on their own if no suitable classes are available. These SATs would also benefit those who are experiencing a situational hearing impairment.

*PhonemeViz* is a SAT we have developed using our framework. It places the character(s) of the most recently spoken initial phoneme just to the side of a speaker's lips. This design should enable a speechreader to focus on the speaker's eyes and lip movements (as in traditional speechreading), while also monitoring changes in PhonemeViz's state using their peripheral vision to help disambiguate confusing visemes. PhonemeViz can be overlaid onto video or displayed on a transparent head mounted display (Figure 4, left) to augment natural speechreading and enhance speechreading acquisition. In a pilot evaluation of a PhonemeViz prototype [9] we found it enabled all participants to achieve 100% word recognition.

When people encounter listening issues due to noise or other factors in their environment this constitutes a situational hearing impairment (e.g., in a noisy restaurant or on a construction site). Through the use of SATs, such as PhonemeViz, people who experience situational hearing impairment can then rely on SR in these challenging situations to facilitate communication.

## Conclusion

With our case studies, we have shown how we can develop perceptual augmentations from research on assistive systems that target disabilities. From research on impaired colour vision, we have arrived at techniques that allow the perception of colour and information outside of the capabilities of the human visual system. From looking at situational visual impairments, we will develop tools and guidelines that can make content on mobile devices easier to perceive. Finally, from investigating hearing impairment, we have arrived at tools that can help people in situations that can cause situational hearing impairment, e.g., noisy environments. While initially each case study focuses on a specific accessibility problem, the results can be generalised and applied to the benefit of everyone.